\documentclass[%
 reprint,
 longbibliography,
 amsmath,amssymb,
 aps,%superscriptaddress
]{revtex4-2}
\usepackage{graphicx}% Include figure files
\usepackage{dcolumn}% Align table columns on decimal point
\usepackage{bm}% bold math
\usepackage[colorlinks=true, linkcolor=blue, citecolor=blue, urlcolor=blue]{hyperref}% add hypertext capabilities
\begin{document}

\preprint{APS/123-QED}

\title{Closed-loop dual-channel atomic beam interferometry beyond the half-fringe limit}% Force line breaks with \\

\author{Wei-Chen Jia \(^{1,2}\)}
\author{Yue Xin \(^{1,2}\)}
\author{Ke Shen \(^{1,2}\)}
\author{Zhi-Xin Meng \(^{3}\)}
\author{Xiang-Xiang Lu \(^{4}\)}
\author{Yi-Cheng Deng \(^{3}\)}
\author{Yuan-Xing Liu \(^{5,6}\)}
\author{Yan-Ying Feng \(^{1,2,}\)} 
\email{yyfeng@tsinghua.edu.cn}

\affiliation{
\(^{1}\) A-Knows Lab, Department of Precision Instruments, Tsinghua University, Beijing 100084, China}

\affiliation{
\(^{2}\) State Key Laboratory of Precision Measurement Technology and Instruments, Beijing 100084, China}

\affiliation{
\(^{3}\) Beijing Institute of Aerospace Control Devices, Beijing 100854, China}

\affiliation{
\(^{4}\) Tianjin Key Laboratory of Quantum Precision Measurement Technology, Tianjin 300131, China}

\affiliation{
\(^{5}\) State Key Laboratory of Ocean Sensing, Zhejiang University, Hangzhou 310058, China}

\affiliation{
\(^{6}\) College of Optical Science and Engineering, Zhejiang University, Hangzhou 310027, China}

\date{\today}

\begin{abstract}
Atom interferometric inertial sensors offer exceptional sensitivity but are fundamentally constrained by the periodic phase response of matter-wave interference, which imposes an intrinsic half-fringe dynamic-range limit and prevents continuous inertial tracking. In multi-axis configurations, additional cross coupling between acceleration and rotation further complicates closed-loop operation. Here we demonstrate the first dual-channel closed-loop operation of an atomic beam interferometer, realizing decoupled feedback control of acceleration- and rotation-induced phases and overcoming the half-fringe limitation. Using continuous, transversely cooled $^{87}$Rb atomic beams, the interferometric phases associated with rotation and acceleration are independently extracted, tracked across multiple fringes, and actively compensated through Raman frequency modulation. This closed-loop scheme enables unambiguous measurements up to $\pm1\,\mathrm{^{\circ}/s}$ in rotation and $\pm0.17\,\mathrm{g}$ in acceleration while maintaining high fringe contrast, corresponding to nearly two orders-of-magnitude extension beyond the conventional half-fringe limit. The sensor achieves a long-term stability of $4\times10^{-4}\,\mathrm{^{\circ}/h}$ for rotation and $4\,\mathrm{\mu g}$ for acceleration at an averaging time of $1000\,\mathrm{s}$. By converting the intrinsically periodic interferometric response into stabilized phase-encoded inertial channels, this work establishes a new operating regime for atomic beam interferometry and advances matter-wave sensors toward practical quantum inertial navigation under dynamic conditions. 

\begin{description}
\item[DOI]
Secondary publications and information retrieval purposes.
\end{description}
\end{abstract}
\maketitle

Atomic interferometers provide a fundamentally stable platform for precision inertial sensing, in which inertial forces are encoded in the accumulated quantum phase of matter waves. Their remarkable sensitivity has enabled applications ranging from tests of fundamental physics \cite{rosi2014precision,hamiltonDark2015a,zhou2015test} to quantum inertial navigation \cite{gustavson1997precision,durfee2006long,canuel2006six,dutta2016continuous,savoieInterleaved2018,gautierAccurate2022,templierTracking2022}. Despite this sensitivity, the interferometric output is intrinsically periodic in phase, imposing a half-fringe dynamic-range limit of $\pm\pi/2$ that restricts unambiguous inertial tracking \cite{bongs2019taking}. In addition, longitudinal velocity dispersion produces phase averaging that degrades fringe contrast and limits robustness under dynamic conditions \cite{sato2025closed}. These intrinsic constraints prevent atom interferometers from operating as continuously stabilized inertial references.

Several strategies have been explored to mitigate these limitations. Hybridization with classical inertial sensors extends the measurable range but introduces additional noise sources and systematic errors \cite{cheiney2018navigation,lautier2014hybridizing,merlet2009operating,jekeli2005navigation}. Reducing interrogation time or interferometer area enlarges the dynamic range at the expense of sensitivity \cite{narducci2022advances}. Interferometer engineering approaches—including multi-species operation \cite{bonnin2018a}, composite fringes \cite{avinadav2020,yankelev2020a}, and velocity modulation \cite{black2023velocity}—increase experimental complexity or compromise bandwidth and data rate. Moreover, these methods cannot mitigate the dephasing arising from a broad atomic velocity distribution, which is particularly severe in atomic beam interferometers. Velocity engineering and three-dimensional cooling \cite{Meng2024,xue2015continuous,kwolek2020three,kwolek2022continuous} alleviate the resulting phase dispersion only at the expense of atomic flux 
and system complexity. None of these approaches fundamentally removes the intrinsic phase periodicity that ultimately limits continuous inertial tracking.

Closed-loop control provides a fundamentally different route. By actively compensating inertial phase shifts and locking the interferometer to its linear response region, closed-loop operation eliminates half-fringe ambiguity while suppressing phase dispersion. %\cite{cahill1979phase,davis1981closed}. 
Continuous atomic beams are particularly advantageous for closed-loop inertial sensing because they provide intrinsically high data rates and near-continuous sampling. This minimizes dead-time-induced phase aliasing, increases the achievable loop bandwidth, and enables robust multi-fringe phase tracking under dynamic inertial inputs. Initial demonstrations have validated single-axis implementations \cite{sato2025closed,Meng2024,d2024atom}. However, a unified architecture enabling simultaneous and decoupled closed-loop control of both rotation and acceleration—essential for practical inertial navigation—remains absent.

\begin{figure*}
    \centering
    \includegraphics[width=1\linewidth]{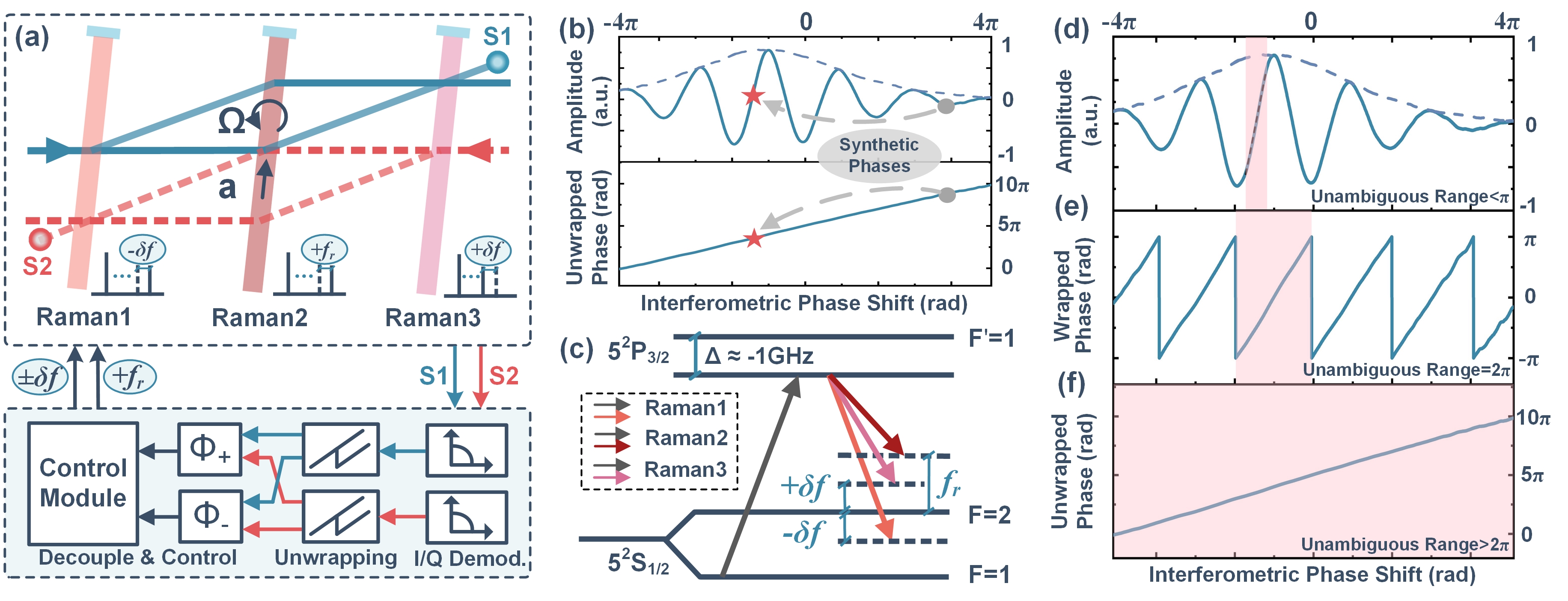}
    \caption{Concept of closed-loop atom-interferometric inertial sensing beyond the conventional half-fringe limitation.
(a) Closed-loop architecture of a Raman Mach–Zehnder (MZ) atom-interferometer inertial sensor enabling simultaneous and decoupled feedback control of acceleration and rotation.
(b) In closed-loop operation, the interferometric phase is locked at the mid-fringe point, avoiding the intrinsic half-fringe ambiguity and suppressing fringe-decay effects, thereby extending the usable dynamic range.
(c) Independent control of acceleration- and rotation-induced phase shifts realized through modulation of the Raman detunings.
(d–f) Evolution of open-loop interferometric readout and its unambiguous range.
(d) Conventional open-loop readout exhibits an intrinsic half-fringe limitation due to the sinusoidal phase dependence of the interference signal.
(e) Quadrature demodulation enables direct phase retrieval and provides a linear response over a full interference period.
(f) Phase unwrapping allows operation across multiple interference periods, further extending the open-loop unambiguous range, but remains susceptible to fringe decay caused by velocity dispersion. Shaded regions in (d)–(f) indicate the corresponding unambiguous measurement ranges.
}
\label{fig:Schematic}
\end{figure*}

Here we demonstrate the first decoupled dual-channel closed-loop operation in an atomic beam interferometer, achieving nearly two orders-of-magnitude enhancement in the dynamic ranges of both acceleration and rotation relative to their intrinsic half-fringe limits. Despite this substantial extension, the sensor preserves long-term stability, reaching $4\times10^{-4}\,^{\circ}/\mathrm{h}$ for rotation and $4\,\mu\mathrm{g}$ for acceleration at an averaging time of $1000\,\mathrm{s}$. Unlike previous implementations, the present work unifies decoupled dual-channel feedback control and phase-based linear readout within a fully quantum architecture, thereby establishing a closed-loop operating paradigm for atomic beam interferometers that combines large dynamic range with long-term stability.

Two-photon stimulated Raman transitions \cite{kasevich1991atomic} act as coherent beam splitters and mirrors for matter waves, coupling $^{87}$Rb atoms between the hyperfine ground states $\lvert 5^2\mathrm{S}_{1/2},F=1 \rangle$ and $\lvert 5^2\mathrm{S}_{1/2},F=2 \rangle$ while imparting momentum $\hbar k_{\mathrm{eff}}$. A $\pi/2-\pi-\pi/2$ pulse sequence realizes a Raman Mach–Zehnder interferometer in which the atoms sample the optical phases of three Raman interactions, producing an interferometric phase determined by inertial and laser-induced contributions, as illustrated in Fig.~\ref{fig:Schematic}(a). 

To overcome the dynamic-range limitation, a closed-loop scheme is implemented in which synthetic phase shifts compensate the inertial phase and lock the interferometer at its optimal operating point [Fig.~\ref{fig:Schematic}(b)]. The synthetic phase is introduced directly into the coherent evolution by modulating the two-photon Raman detunings. Specifically, opposite detunings $\delta_{1,3}=\mp\delta f$ are applied to the two $\pi/2$ pulses, while an independent detuning $\delta_2=f_r$ is applied to the $\pi$ pulse [Fig.~\ref{fig:Schematic}(c)]. These detunings modify the laser phases according to $\Delta\phi_{0,i}=2\pi\delta_i t$, generating a composite phase shift $\Delta \phi_{0,1}-2\Delta \phi_{0,2}+\Delta \phi_{0,3}$, which provides a tunable synthetic compensation of the inertial phase accumulation. This configuration ensures that the synthetic phase enters the two interferometers with opposite signs, enabling independent control of the acceleration and rotation channels.

To generate synthetic phases that independently address rotation and acceleration, a dual Mach–Zehnder configuration is employed. Owing to the symmetric geometry, acceleration and rotation are encoded into the average and half-difference of the two interferometric phases, enabling composite and decoupled closed-loop control of the two inertial channels [Fig.~\ref{fig:Schematic}(a)]. The two interference signals can be expressed as \cite{Detail1}
\begin{equation}
\begin{aligned}
S_1 ={}& A_1 + \frac{C_1}{2}
\cos\Big[
    k_{\mathrm{eff}} a T^2
    - 2 k_{\mathrm{eff}} \Omega L T
    - 4\pi f_r (t_0 + T)  \\
&\qquad\quad
    + 4\pi \delta f T
    + \phi_0
\Big], \\
S_2 ={}& A_2 + \frac{C_2}{2}
\cos\Big[
    - k_{\mathrm{eff}} a T^2
    - 2 k_{\mathrm{eff}} \Omega L T
    - 4\pi f_r (t_0 + T) \\
&\qquad\quad
    - 4\pi \delta f T
    + \phi_0'
\Big],\label{eq:signal}
\end{aligned}
\end{equation}
Here $T=L/v$ denotes the interrogation time between adjacent Raman interactions, with $L$ the spatial separation and $v$ the atomic velocity. The parameter $t_0$ represents the initial time of the interferometric sequence, while $\phi_0$ and $\phi_0'$ denote the initial laser phases of the two interferometers. These expressions reveal that inertial and synthetic phases enter the two interferometers with opposite symmetry, enabling decoupling through sum and difference combinations. 

\begin{figure*}
    \centering
    \includegraphics[width=1\linewidth]{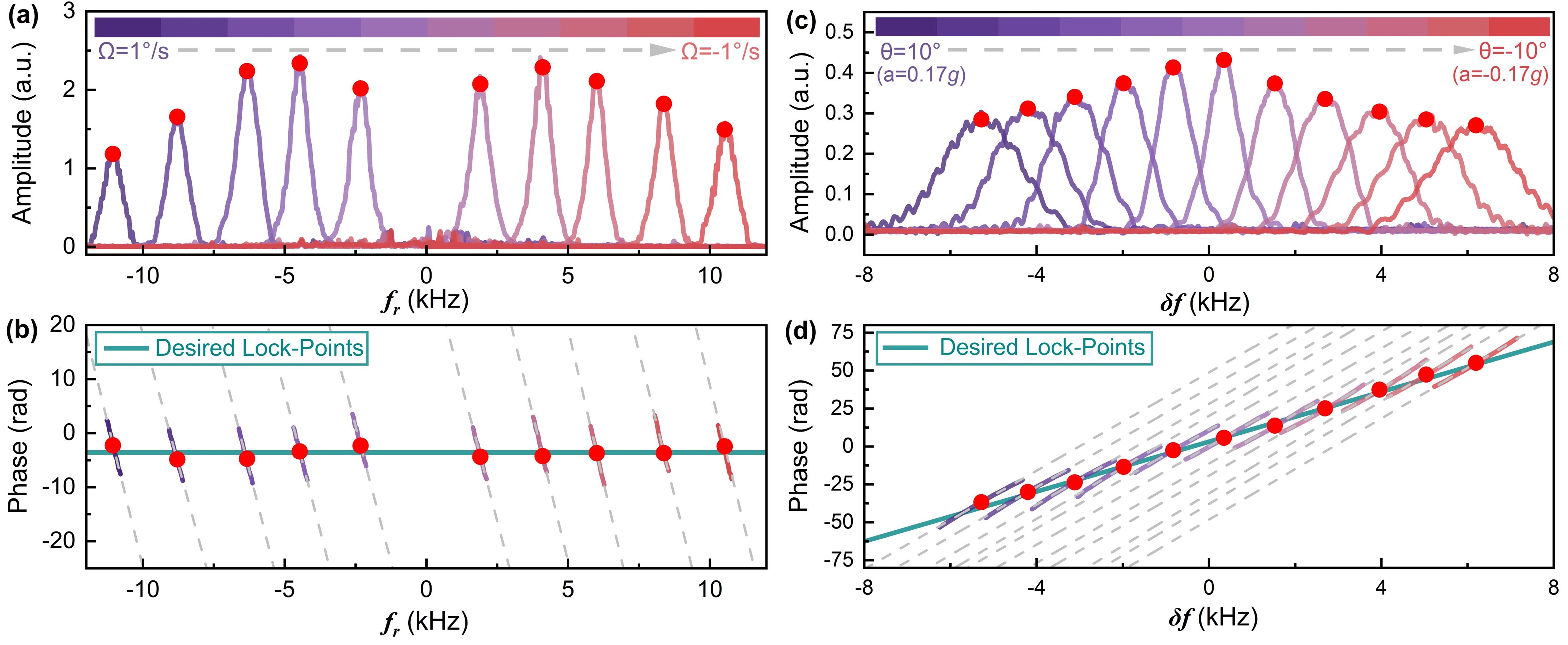}
    \caption{Phase control demonstration for dual-channel closed-loop operation. (a,b) Interferometric amplitude (a) and extracted phase of the left interferomter (b) versus Raman detuning $f_r$ under varying rotation inputs. (c,d) Corresponding amplitude (c) and phase (d) versus $\delta f$ under different acceleration inputs. %$\Phi_{\Omega,cl}$ and $\Phi_{a,cl}$ denote the required phase lock-points derived in Eq.~(\ref{eq:CL-phi}). 
    Quadrature demodulation provides simultaneous amplitude and linear phase readout.}
    \label{fig:Principle}
\end{figure*}

Denoting the two interferometric phases as $\Phi_1$ and $\Phi_2$, the inertial coupling can be written compactly as
\begin{equation}
\boldsymbol{\Phi}
=
\mathbf{M}
\begin{pmatrix}
a \\
\Omega
\end{pmatrix}
+
\mathbf{B} u
+
\boldsymbol{\Phi}_0 ,\label{eq:matrix}
\end{equation}
where $\mathbf{M}$ denotes the inertial coupling matrix and $\mathbf{u}=(4\pi\delta f T, -4\pi f_r T)^T$ are the synthetic phase control parameters. Transforming to the sum and difference basis,
$\Phi_{\pm}=(\Phi_1\pm\Phi_2)/2$, diagonalizes the coupling,
\begin{equation}
\begin{pmatrix}
\Phi_{-} \\
\Phi_{+}
\end{pmatrix}
=
\begin{pmatrix}
k_{\mathrm{eff}} T^2 & 0 \\
0 & -2 k_{\mathrm{eff}} L T
\end{pmatrix}
\begin{pmatrix}
a \\
\Omega
\end{pmatrix}
+4\pi T
\begin{pmatrix}
\delta f \\
-f_r
\end{pmatrix}+\boldsymbol{\Phi}_{0,\mp},\label{eq:decouple}
\end{equation}
explicitly demonstrating decoupled acceleration and rotation channels.

The interferometric phase is extracted by encoding the signal as a phase-modulated carrier at frequency $2f_r$ and performing quadrature demodulation at $4\pi f_r$. Using a digital Hilbert transform to obtain the analytic signal enables direct phase retrieval, yielding an intrinsically linear phase response over $\pm\pi$ in open-loop operation without fringe-amplitude calibration. Conventional interferometric readout measures only the cosine projection of the interferometric phase, leading to an intrinsic half-fringe ambiguity. Quadrature detection reconstructs the full complex interferometric phasor, enabling direct phase extraction over a full $2\pi$ range. Multi-fringe tracking is realized through phase unwrapping, correcting inter-cycle $2\pi$ discontinuities under high–data-rate conditions. The interferometer therefore operates as a phase-encoded quantum sensor rather than a fringe-amplitude detector.

The phase-demodulation scheme and the resulting extension of the open-loop dynamic range are experimentally demonstrated in Fig.~\ref{fig:Schematic}(d-f). The interferometric phase is scanned by modulating the detuning $\delta f$, which introduces a phase shift $\phi=4\pi \delta f t$. The demodulated fringe amplitude obtained through quadrature detection is shown in Fig.~\ref{fig:Schematic}(d), exhibiting the conventional sinusoidal interferometric response with a decaying envelope. In this representation the measurement becomes ambiguous outside the monotonic region around the fringe center, corresponding to the conventional half-fringe unambiguous range. By reconstructing the analytic signal, the interferometric phase can instead be directly retrieved through quadrature demodulation, as shown in Fig.~\ref{fig:Schematic}(e), yielding a linear phase response around the mid-fringe while providing wrapped phase readout over the interval $[-\pi,\pi]$, without relying on fringe-amplitude calibration. Applying phase unwrapping further enables continuous tracking across successive $2\pi$ phase cycles, producing the unwrapped phase evolution shown in Fig.~\ref{fig:Schematic}(f). In our experiment the open-loop phase-tracking range extends to approximately $\pm 4\pi$. This phase-encoding and unwrapping technique significantly enlarges the measurable range compared with amplitude readout, preventing erroneous closed-loop locking and enabling a more robust response to rapidly varying inertial inputs. However, the approach remains limited by contrast reduction arising from velocity-distribution–induced dephasing, motivating the closed-loop phase-compensation scheme described below.

\begin{figure*}
    \centering
    \includegraphics[width=1\linewidth]{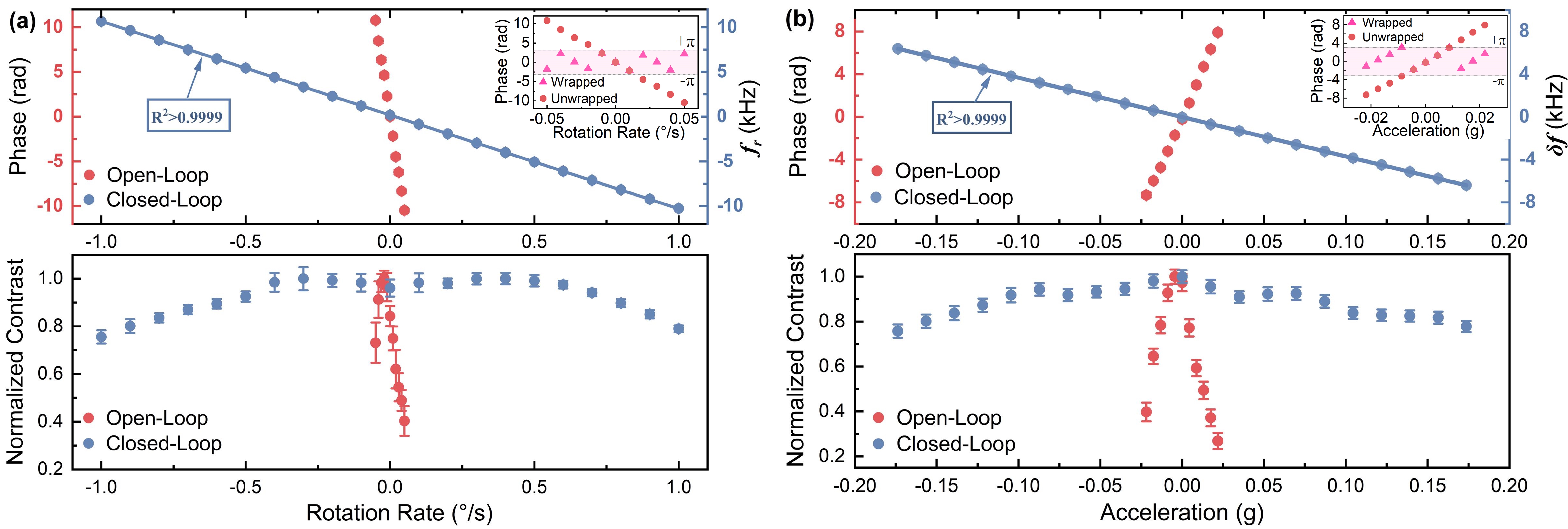}
    \caption{Dynamic-range extension under closed-loop operation. (a) Rotation calibration comparing open-loop (phase output) and closed-loop (frequency-encoded output) operation. Lower panel: corresponding normalized fringe contrast. (b) Acceleration calibration (upper) and associated fringe contrast (lower). Inset: comparison of open-loop operation with and without phase unwrapping.}
    \label{fig:dynamic}
\end{figure*}

To preserve fringe contrast in the presence of longitudinal velocity spread, the closed-loop controller enforces first-order velocity insensitivity of the locked phase. Imposing $\partial\Phi_+/\partial v=0$ and $\partial\Phi_-/\partial v=0$, the locked-point can be derived as
\begin{equation}
\Phi_{\Omega,cl} =\Phi_{0,+}, 
\qquad
\Phi_{a,cl} =2\pi (L/v_{eq})\delta f+\Phi_{0,-},
\label{eq:CL-phi}
\end{equation}
where $v_{\mathrm{eq}}$ characterizes the effective ensemble velocity. It is emphasized that $\Phi_{a,cl}$ is not constant but depends linearly with $\delta f$, owing to the distinct velocity scaling in the acceleration-induced phase shift $\phi_a=k_{\mathrm{eff}}aT^2$ and the synthetic phase term $\phi_s=4\pi \delta fT$.
The stabilization of rotation and acceleration channels leads to
\begin{equation}
\Omega = -\frac{2\pi f_r}{k_{\mathrm{eff}} L}, 
\qquad
a = -\frac{2\pi v_{\mathrm{eq}}\delta f}{k_{\mathrm{eff}} L},\label{eq:CL}
\end{equation}
In this regime, the interferometric phase is stabilized while the inertial information is transformed into frequency-control parameters, enabling extended dynamic range with preserved precision.

The experiment employs two counter-propagating, transversely cooled $^{87}$Rb atomic beams with a most probable longitudinal velocity of $\sim200\,\mathrm{m/s}$. Three pairs of retro-reflected Raman beams separated by $2L=540\,\mathrm{mm}$ form dual Mach–Zehnder interferometers \cite{Yan2025}. Independent control of the Raman two-photon detunings is achieved using a multi-channel microwave synthesizer driving electro-optic modulators. The interferometric signals $S_1$ and $S_2$ are detected via laser-induced fluorescence. Phase extraction, Hilbert-based quadrature demodulation, phase unwrapping, and closed-loop feedback are implemented digitally. The control parameters $f_r$ and $\delta f$ are updated at 200~Hz via independent PID controllers. Rotation is locked to $\Phi_{\mathrm{set1}}=0$, while acceleration follows the linear locking condition in Eq.~(\ref{eq:CL-phi}). Inertial calibration is performed using a precision turntable and a tilt platform that projects gravity onto the Raman wave-vector direction.

The phase-control behavior is experimentally investigated in Fig.~\ref{fig:Principle}. For rotation sensing, open-loop modulation shows that the interferometric phase can be maintained at a constant set point while the control parameter $f_r$ varies linearly with the applied rotation, thereby preserving maximal fringe contrast [Fig.~\ref{fig:Principle}(a,b)]. Similarly, for acceleration sensing, the optimal lock points exhibit a linear dependence on $\delta f$, consistent with Eq.~(\ref{eq:CL-phi}) [Fig.~\ref{fig:Principle}(c,d)]. In both channels, amplitude and phase are simultaneously retrieved via quadrature demodulation, confirming linear phase readout over an extended range.

Figure~\ref{fig:dynamic} quantifies the dynamic-range enhancement enabled by closed-loop operation. In open-loop mode, the measurable range is restricted to the intrinsic half-fringe interval. Under closed-loop operation, the inertial signal is encoded directly in the control parameters $f_r$ and $\delta f$, enabling continuous tracking across multiple fringes while maintaining high contrast. For rotation [Fig.~\ref{fig:dynamic}(a)], the dynamic range extends to $\pm1\,^{\circ}/\mathrm{s}$, corresponding to nearly two orders-of-magnitude larger inertial dynamic range relative to the intrinsic half-fringe dynamic range, with negligible contrast degradation. For acceleration [Fig.~\ref{fig:dynamic}(b)], the sensor achieves a range of $\pm0.17\,\mathrm{g}$ while preserving interferometric visibility. The inset demonstrates that phase unwrapping alone extends the open-loop range but does not prevent contrast loss, highlighting the essential role of closed-loop compensation.

We analyze the deviation of the measured scale factor from its nominal value. For rotation sensing, the expected scale factor is $-k_{\mathrm{eff}}L/(2\pi)=-12082~\mathrm{Hz/(^{\circ}/s)}$, whereas the experimentally extracted value is 
$-10473~\mathrm{Hz/(^{\circ}/s)}$, corresponding to a relative deviation of $13.32\%$.
This discrepancy is primarily attributed to velocity-selective effects inherent to the atomic beam configuration. The effective longitudinal velocities participating in real rotation measurements ($v_{eq,r}$) and in synthetic rotation induced via $f_r$ modulation ($v_{eq,f}$) are calibrated as $v_{eq,r}=(190.64\pm0.65)\,\mathrm{m/s}$ and $v_{eq,f}=(165.72\pm0.52)\,\mathrm{m/s}$, respectively. Their relative difference of $13.16\%$ closely matches the observed scale-factor deviation, indicating that imperfect velocity equivalence between physical and synthetic phase compensation dominates the systematic error. Further improvement is expected through narrowing or actively selecting the longitudinal velocity distribution, thereby reducing velocity-dependent weighting in the closed-loop response. This analysis highlights the central role of velocity distribution in atomic-beam closed-loop interferometry.

Similar to the envelope reduction observed in Fig.~\ref{fig:Principle}, a gradual decrease of fringe contrast persists with increasing inertial input even under closed-loop operation. The fringe contrast can be written as
\begin{equation}
C=\left|\int P(v)\,e^{i\Phi(v)}\,\mathrm{d}v\right|,
\end{equation}
where $\Phi(v)$ denotes the velocity-dependent interferometric phase and $P(v)$ is the longitudinal velocity distribution.

Expanding $\Phi(v)$ around the central velocity $v_0$,
\begin{equation}
\begin{aligned}
\Phi(v)=\;&\Phi(v_0)
+\left.\frac{\partial \Phi}{\partial v}\right|_{v_0}(v-v_0)\\
&+\frac{1}{2}\left.\frac{\partial^2 \Phi}{\partial v^2}\right|_{v_0}(v-v_0)^2
+\cdots,
\end{aligned}
\end{equation}
reveals that contrast reduction arises from residual velocity-dependent phase dispersion.

\begin{figure}
    \centering
    \includegraphics[width=0.98\linewidth]{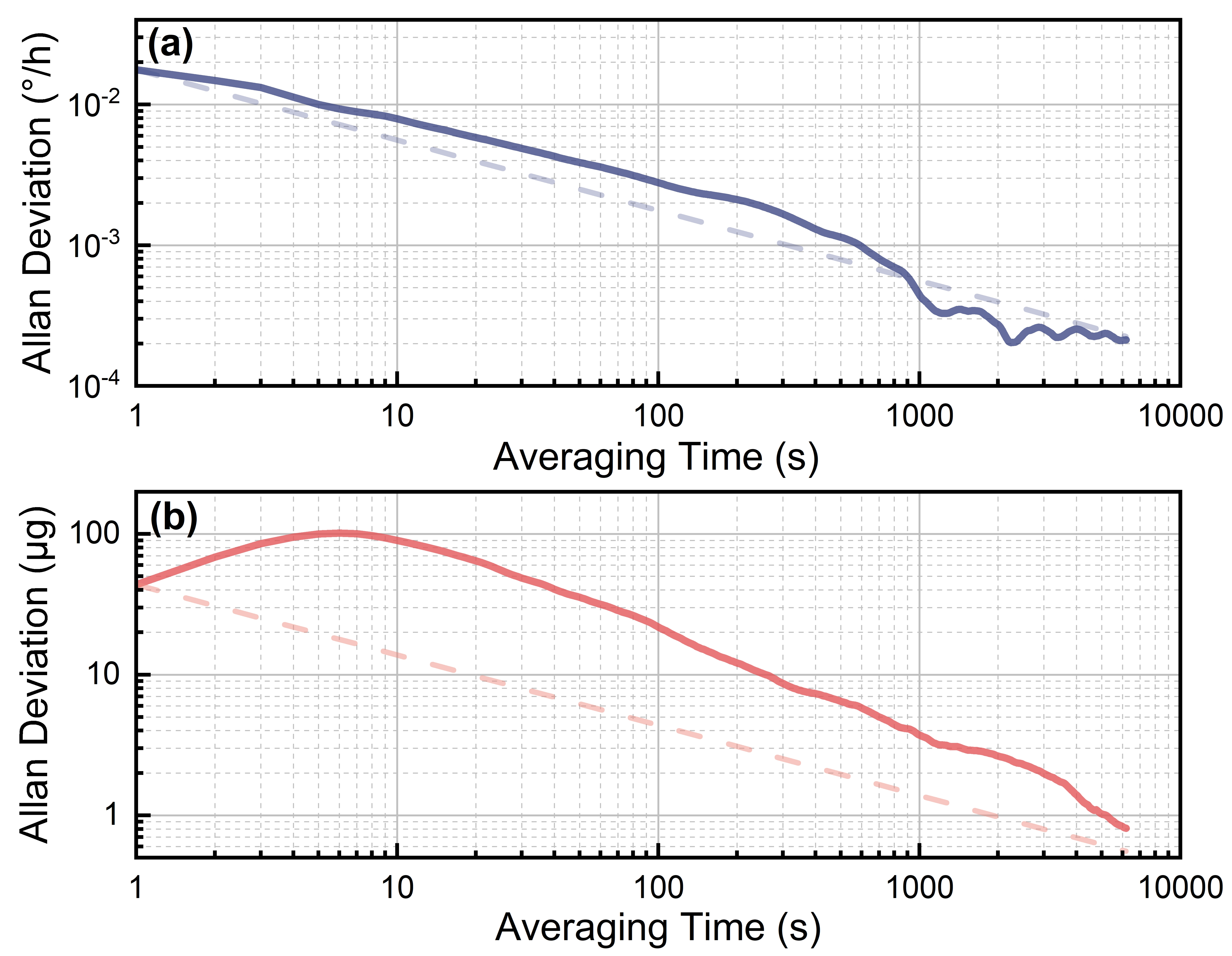}
    \caption{Long-term stability of the dual-axis inertial sensor. Allan deviation of rotation (a) and acceleration (b) measured over 25,000~s. Dashed lines indicate a $1/\sqrt{\tau}$ dependence, where $\tau$ is the averaging time.}
    \label{fig:allan}
\end{figure}

For rotation sensing, imperfect equivalence between $v_{eq,r}$ and $v_{eq,f}$ prevents complete cancellation of velocity dependence, leading to non-zero higher-order derivatives $\partial^n\Phi_+/\partial v^n$ and accelerated contrast decay \cite{Detail2}. In contrast, for acceleration sensing the closed-loop condition enforces $\partial\Phi_{-}/\partial v=0$, eliminating the first-order velocity-dependent phase term. The remaining higher-order contributions result in a slower residual decay of fringe contrast at large inertial inputs. This behavior highlights that closed-loop operation suppresses first-order velocity dispersion but does not eliminate higher-order dephasing inherent to finite velocity spread.

Beyond the substantial extension of dynamic range, the sensor simultaneously achieves long-term bias stability. The inertial phase is extracted via quadrature demodulation with high signal-to-noise ratio, while technical noise and drift sources—including laser phase noise, radio-frequency fluctuations, and environmental perturbations—are minimized. Additional temperature compensation suppress residual systematic drifts. As shown in Fig.~\ref{fig:allan}, the Allan deviation stability reaches $4\times10^{-4}\,\mathrm{^{\circ}/h}$ for rotation and $4\,\mathrm{\mu g}$ for acceleration at an averaging time of $1000\,\mathrm{s}$, demonstrating stable dual-channel operation over 25,000~s. 

Further improvements could be obtained by enhancing the efficiency of Raman coherent manipulation to increase interferometric contrast. Since the short-term sensitivity scales inversely with fringe contrast, higher-contrast operation would directly improve the signal-to-noise ratio. Under realistic contrast enhancement and noise suppression, an order-of-magnitude improvement in bias stability appears feasible, potentially placing atomic beam interferometric gyroscopes in the $10^{-5}\,\mathrm{^{\circ}/h}$ regime even for a reduced interferometer length \cite{durfee2006long}. These prospects highlight the scalability of the closed-loop atomic beam architecture for compact high-performance inertial sensors.

In conclusion, we demonstrate the first atomic interferometric inertial sensor operating in simultaneous and decoupled closed-loop control of both rotation and acceleration. By stabilizing the interferometric phases against intrinsic phase periodicity and velocity-induced dephasing, the sensor extends the usable dynamic range by nearly two orders of magnitude beyond the conventional $[-\pi/2,\pi/2]$ linear region while preserving stability. The fully quantum architecture supports high data-update rates without auxiliary inertial sensors or complex measurement sequences, enabling stable dual-channel operation under dynamic conditions.

More broadly, the present work establishes a new operating regime for atomic beam interferometry in which inertial information is encoded directly in control parameters rather than fringe amplitude. By transforming atom interferometers from fringe-amplitude sensors into stabilized phase-encoded quantum inertial sensors, this approach advances matter-wave interferometry toward practical inertial reference systems capable of operating under realistic dynamic conditions.

\bibliography{apssamp}% Produces the bibliography via BibTeX.

\providecommand{\noopsort}[1]{}\providecommand{\singleletter}[1]{#1}%
\begin{thebibliography}{30}%
\makeatletter
\providecommand \@ifxundefined [1]{%
 \@ifx{#1\undefined}
}%
\providecommand \@ifnum [1]{%
 \ifnum #1\expandafter \@firstoftwo
 \else \expandafter \@secondoftwo
 \fi
}%
\providecommand \@ifx [1]{%
 \ifx #1\expandafter \@firstoftwo
 \else \expandafter \@secondoftwo
 \fi
}%
\providecommand \natexlab [1]{#1}%
\providecommand \enquote  [1]{``#1''}%
\providecommand \bibnamefont  [1]{#1}%
\providecommand \bibfnamefont [1]{#1}%
\providecommand \citenamefont [1]{#1}%
\providecommand \href@noop [0]{\@secondoftwo}%
\providecommand \href [0]{\begingroup \@sanitize@url \@href}%
\providecommand \@href[1]{\@@startlink{#1}\@@href}%
\providecommand \@@href[1]{\endgroup#1\@@endlink}%
\providecommand \@sanitize@url [0]{\catcode `\\12\catcode `\$12\catcode
  `\&12\catcode `\#12\catcode `\^12\catcode `\_12\catcode `\%12\relax}%
\providecommand \@@startlink[1]{}%
\providecommand \@@endlink[0]{}%
\providecommand \url  [0]{\begingroup\@sanitize@url \@url }%
\providecommand \@url [1]{\endgroup\@href {#1}{\urlprefix }}%
\providecommand \urlprefix  [0]{URL }%
\providecommand \Eprint [0]{\href }%
\providecommand \doibase [0]{https://doi.org/}%
\providecommand \selectlanguage [0]{\@gobble}%
\providecommand \bibinfo  [0]{\@secondoftwo}%
\providecommand \bibfield  [0]{\@secondoftwo}%
\providecommand \translation [1]{[#1]}%
\providecommand \BibitemOpen [0]{}%
\providecommand \bibitemStop [0]{}%
\providecommand \bibitemNoStop [0]{.\EOS\space}%
\providecommand \EOS [0]{\spacefactor3000\relax}%
\providecommand \BibitemShut  [1]{\csname bibitem#1\endcsname}%
\let\auto@bib@innerbib\@empty
%</preamble>
\bibitem [{\citenamefont {Rosi}\ \emph {et~al.}(2014)\citenamefont {Rosi},
  \citenamefont {Sorrentino}, \citenamefont {Cacciapuoti}, \citenamefont
  {Prevedelli},\ and\ \citenamefont {Tino}}]{rosi2014precision}%
  \BibitemOpen
  \bibfield  {author} {\bibinfo {author} {\bibfnamefont {G.}~\bibnamefont
  {Rosi}}, \bibinfo {author} {\bibfnamefont {F.}~\bibnamefont {Sorrentino}},
  \bibinfo {author} {\bibfnamefont {L.}~\bibnamefont {Cacciapuoti}}, \bibinfo
  {author} {\bibfnamefont {M.}~\bibnamefont {Prevedelli}},\ and\ \bibinfo
  {author} {\bibfnamefont {G.}~\bibnamefont {Tino}},\ }\bibfield  {title}
  {\bibinfo {title} {Precision measurement of the newtonian gravitational
  constant using cold atoms},\ }\href
  {https://doi.org/https://doi.org/10.1038/nature13433} {\bibfield  {journal}
  {\bibinfo  {journal} {Nature}\ }\textbf {\bibinfo {volume} {510}},\ \bibinfo
  {pages} {518} (\bibinfo {year} {2014})}\BibitemShut {NoStop}%
\bibitem [{\citenamefont {Hamilton}\ \emph {et~al.}(2015)\citenamefont
  {Hamilton}, \citenamefont {Jaffe}, \citenamefont {Haslinger}, \citenamefont
  {Simmons}, \citenamefont {M{\"u}ller},\ and\ \citenamefont
  {Khoury}}]{hamiltonDark2015a}%
  \BibitemOpen
  \bibfield  {author} {\bibinfo {author} {\bibfnamefont {P.}~\bibnamefont
  {Hamilton}}, \bibinfo {author} {\bibfnamefont {M.}~\bibnamefont {Jaffe}},
  \bibinfo {author} {\bibfnamefont {P.}~\bibnamefont {Haslinger}}, \bibinfo
  {author} {\bibfnamefont {Q.}~\bibnamefont {Simmons}}, \bibinfo {author}
  {\bibfnamefont {H.}~\bibnamefont {M{\"u}ller}},\ and\ \bibinfo {author}
  {\bibfnamefont {J.}~\bibnamefont {Khoury}},\ }\bibfield  {title} {\bibinfo
  {title} {Atom-interferometry constraints on dark energy},\ }\href
  {https://doi.org/10.1126/science.aaa8883} {\bibfield  {journal} {\bibinfo
  {journal} {Science}\ }\textbf {\bibinfo {volume} {349}},\ \bibinfo {pages}
  {849} (\bibinfo {year} {2015})}\BibitemShut {NoStop}%
\bibitem [{\citenamefont {Zhou}\ \emph {et~al.}(2015)\citenamefont {Zhou},
  \citenamefont {Long}, \citenamefont {Tang}, \citenamefont {Chen},
  \citenamefont {Gao}, \citenamefont {Peng}, \citenamefont {Duan},
  \citenamefont {Zhong}, \citenamefont {Xiong}, \citenamefont {Wang} \emph
  {et~al.}}]{zhou2015test}%
  \BibitemOpen
  \bibfield  {author} {\bibinfo {author} {\bibfnamefont {L.}~\bibnamefont
  {Zhou}}, \bibinfo {author} {\bibfnamefont {S.}~\bibnamefont {Long}}, \bibinfo
  {author} {\bibfnamefont {B.}~\bibnamefont {Tang}}, \bibinfo {author}
  {\bibfnamefont {X.}~\bibnamefont {Chen}}, \bibinfo {author} {\bibfnamefont
  {F.}~\bibnamefont {Gao}}, \bibinfo {author} {\bibfnamefont {W.}~\bibnamefont
  {Peng}}, \bibinfo {author} {\bibfnamefont {W.}~\bibnamefont {Duan}}, \bibinfo
  {author} {\bibfnamefont {J.}~\bibnamefont {Zhong}}, \bibinfo {author}
  {\bibfnamefont {Z.}~\bibnamefont {Xiong}}, \bibinfo {author} {\bibfnamefont
  {J.}~\bibnamefont {Wang}}, \emph {et~al.},\ }\bibfield  {title} {\bibinfo
  {title} {Test of equivalence principle at $10^{-8}$ level by a dual-species
  double-diffraction raman atom interferometer},\ }\href
  {https://doi.org/https://doi.org/10.1103/PhysRevLett.115.013004} {\bibfield
  {journal} {\bibinfo  {journal} {Phys. Rev. Lett.}\ }\textbf {\bibinfo
  {volume} {115}},\ \bibinfo {pages} {013004} (\bibinfo {year}
  {2015})}\BibitemShut {NoStop}%
\bibitem [{\citenamefont {Gustavson}\ \emph {et~al.}(1997)\citenamefont
  {Gustavson}, \citenamefont {Bouyer},\ and\ \citenamefont
  {Kasevich}}]{gustavson1997precision}%
  \BibitemOpen
  \bibfield  {author} {\bibinfo {author} {\bibfnamefont {T.}~\bibnamefont
  {Gustavson}}, \bibinfo {author} {\bibfnamefont {P.}~\bibnamefont {Bouyer}},\
  and\ \bibinfo {author} {\bibfnamefont {M.}~\bibnamefont {Kasevich}},\
  }\bibfield  {title} {\bibinfo {title} {Precision rotation measurements with
  an atom interferometer gyroscope},\ }\href
  {https://doi.org/https://doi.org/10.1103/PhysRevLett.78.2046} {\bibfield
  {journal} {\bibinfo  {journal} {Phys. Rev. Lett.}\ }\textbf {\bibinfo
  {volume} {78}},\ \bibinfo {pages} {2046} (\bibinfo {year}
  {1997})}\BibitemShut {NoStop}%
\bibitem [{\citenamefont {Durfee}\ \emph {et~al.}(2006)\citenamefont {Durfee},
  \citenamefont {Shaham},\ and\ \citenamefont {Kasevich}}]{durfee2006long}%
  \BibitemOpen
  \bibfield  {author} {\bibinfo {author} {\bibfnamefont {D.}~\bibnamefont
  {Durfee}}, \bibinfo {author} {\bibfnamefont {Y.}~\bibnamefont {Shaham}},\
  and\ \bibinfo {author} {\bibfnamefont {M.}~\bibnamefont {Kasevich}},\
  }\bibfield  {title} {\bibinfo {title} {Long-term stability of an
  area-reversible atom-interferometer sagnac gyroscope},\ }\href
  {https://doi.org/https://doi.org/10.1103/PhysRevLett.97.240801} {\bibfield
  {journal} {\bibinfo  {journal} {Phys. Rev. Lett.}\ }\textbf {\bibinfo
  {volume} {97}},\ \bibinfo {pages} {240801} (\bibinfo {year}
  {2006})}\BibitemShut {NoStop}%
\bibitem [{\citenamefont {Canuel}\ \emph {et~al.}(2006)\citenamefont {Canuel},
  \citenamefont {Leduc}, \citenamefont {Holleville}, \citenamefont {Gauguet},
  \citenamefont {Fils}, \citenamefont {Virdis}, \citenamefont {Clairon},
  \citenamefont {Dimarcq}, \citenamefont {Bord{\'e}}, \citenamefont {Landragin}
  \emph {et~al.}}]{canuel2006six}%
  \BibitemOpen
  \bibfield  {author} {\bibinfo {author} {\bibfnamefont {B.}~\bibnamefont
  {Canuel}}, \bibinfo {author} {\bibfnamefont {F.}~\bibnamefont {Leduc}},
  \bibinfo {author} {\bibfnamefont {D.}~\bibnamefont {Holleville}}, \bibinfo
  {author} {\bibfnamefont {A.}~\bibnamefont {Gauguet}}, \bibinfo {author}
  {\bibfnamefont {J.}~\bibnamefont {Fils}}, \bibinfo {author} {\bibfnamefont
  {A.}~\bibnamefont {Virdis}}, \bibinfo {author} {\bibfnamefont
  {A.}~\bibnamefont {Clairon}}, \bibinfo {author} {\bibfnamefont
  {N.}~\bibnamefont {Dimarcq}}, \bibinfo {author} {\bibfnamefont {C.~J.}\
  \bibnamefont {Bord{\'e}}}, \bibinfo {author} {\bibfnamefont {A.}~\bibnamefont
  {Landragin}}, \emph {et~al.},\ }\bibfield  {title} {\bibinfo {title}
  {Six-axis inertial sensor using cold-atom interferometry},\ }\href
  {https://doi.org/https://doi.org/10.1103/PhysRevLett.97.010402} {\bibfield
  {journal} {\bibinfo  {journal} {Phys. Rev. Lett.}\ }\textbf {\bibinfo
  {volume} {97}},\ \bibinfo {pages} {010402} (\bibinfo {year}
  {2006})}\BibitemShut {NoStop}%
\bibitem [{\citenamefont {Dutta}\ \emph {et~al.}(2016)\citenamefont {Dutta},
  \citenamefont {Savoie}, \citenamefont {Fang}, \citenamefont {Venon},
  \citenamefont {Alzar}, \citenamefont {Geiger},\ and\ \citenamefont
  {Landragin}}]{dutta2016continuous}%
  \BibitemOpen
  \bibfield  {author} {\bibinfo {author} {\bibfnamefont {I.}~\bibnamefont
  {Dutta}}, \bibinfo {author} {\bibfnamefont {D.}~\bibnamefont {Savoie}},
  \bibinfo {author} {\bibfnamefont {B.}~\bibnamefont {Fang}}, \bibinfo {author}
  {\bibfnamefont {B.}~\bibnamefont {Venon}}, \bibinfo {author} {\bibfnamefont
  {C.~G.}\ \bibnamefont {Alzar}}, \bibinfo {author} {\bibfnamefont
  {R.}~\bibnamefont {Geiger}},\ and\ \bibinfo {author} {\bibfnamefont
  {A.}~\bibnamefont {Landragin}},\ }\bibfield  {title} {\bibinfo {title}
  {Continuous cold-atom inertial sensor with 1 nrad/sec rotation stability},\
  }\href {https://doi.org/https://doi.org/10.1103/PhysRevLett.116.183003}
  {\bibfield  {journal} {\bibinfo  {journal} {Phys. Rev. Lett.}\ }\textbf
  {\bibinfo {volume} {116}},\ \bibinfo {pages} {183003} (\bibinfo {year}
  {2016})}\BibitemShut {NoStop}%
\bibitem [{\citenamefont {Savoie}\ \emph {et~al.}(2018)\citenamefont {Savoie},
  \citenamefont {Altorio}, \citenamefont {Fang}, \citenamefont {Sidorenkov},
  \citenamefont {Geiger},\ and\ \citenamefont
  {Landragin}}]{savoieInterleaved2018}%
  \BibitemOpen
  \bibfield  {author} {\bibinfo {author} {\bibfnamefont {D.}~\bibnamefont
  {Savoie}}, \bibinfo {author} {\bibfnamefont {M.}~\bibnamefont {Altorio}},
  \bibinfo {author} {\bibfnamefont {B.}~\bibnamefont {Fang}}, \bibinfo {author}
  {\bibfnamefont {L.~A.}\ \bibnamefont {Sidorenkov}}, \bibinfo {author}
  {\bibfnamefont {R.}~\bibnamefont {Geiger}},\ and\ \bibinfo {author}
  {\bibfnamefont {A.}~\bibnamefont {Landragin}},\ }\bibfield  {title} {\bibinfo
  {title} {Interleaved atom interferometry for high-sensitivity inertial
  measurements},\ }\href {https://doi.org/10.1126/sciadv.aau7948} {\bibfield
  {journal} {\bibinfo  {journal} {Sci. Adv.}\ }\textbf {\bibinfo {volume}
  {4}},\ \bibinfo {pages} {eaau7948} (\bibinfo {year} {2018})}\BibitemShut
  {NoStop}%
\bibitem [{\citenamefont {Gautier}\ \emph {et~al.}(2022)\citenamefont
  {Gautier}, \citenamefont {Guessoum}, \citenamefont {Sidorenkov},
  \citenamefont {Bouton}, \citenamefont {Landragin},\ and\ \citenamefont
  {Geiger}}]{gautierAccurate2022}%
  \BibitemOpen
  \bibfield  {author} {\bibinfo {author} {\bibfnamefont {R.}~\bibnamefont
  {Gautier}}, \bibinfo {author} {\bibfnamefont {M.}~\bibnamefont {Guessoum}},
  \bibinfo {author} {\bibfnamefont {L.~A.}\ \bibnamefont {Sidorenkov}},
  \bibinfo {author} {\bibfnamefont {Q.}~\bibnamefont {Bouton}}, \bibinfo
  {author} {\bibfnamefont {A.}~\bibnamefont {Landragin}},\ and\ \bibinfo
  {author} {\bibfnamefont {R.}~\bibnamefont {Geiger}},\ }\bibfield  {title}
  {\bibinfo {title} {Accurate measurement of the {{Sagnac}} effect for matter
  waves},\ }\href {https://doi.org/10.1126/sciadv.abn8009} {\bibfield
  {journal} {\bibinfo  {journal} {Sci. Adv.}\ }\textbf {\bibinfo {volume}
  {8}},\ \bibinfo {pages} {eabn8009} (\bibinfo {year} {2022})}\BibitemShut
  {NoStop}%
\bibitem [{\citenamefont {Templier}\ \emph {et~al.}(2022)\citenamefont
  {Templier}, \citenamefont {Cheiney}, \citenamefont {{d'Armagnac de
  Castanet}}, \citenamefont {Gouraud}, \citenamefont {Porte}, \citenamefont
  {Napolitano}, \citenamefont {Bouyer}, \citenamefont {Battelier},\ and\
  \citenamefont {Barrett}}]{templierTracking2022}%
  \BibitemOpen
  \bibfield  {author} {\bibinfo {author} {\bibfnamefont {S.}~\bibnamefont
  {Templier}}, \bibinfo {author} {\bibfnamefont {P.}~\bibnamefont {Cheiney}},
  \bibinfo {author} {\bibfnamefont {Q.}~\bibnamefont {{d'Armagnac de
  Castanet}}}, \bibinfo {author} {\bibfnamefont {B.}~\bibnamefont {Gouraud}},
  \bibinfo {author} {\bibfnamefont {H.}~\bibnamefont {Porte}}, \bibinfo
  {author} {\bibfnamefont {F.}~\bibnamefont {Napolitano}}, \bibinfo {author}
  {\bibfnamefont {P.}~\bibnamefont {Bouyer}}, \bibinfo {author} {\bibfnamefont
  {B.}~\bibnamefont {Battelier}},\ and\ \bibinfo {author} {\bibfnamefont
  {B.}~\bibnamefont {Barrett}},\ }\bibfield  {title} {\bibinfo {title}
  {Tracking the vector acceleration with a hybrid quantum accelerometer
  triad},\ }\href {https://doi.org/10.1126/sciadv.add3854} {\bibfield
  {journal} {\bibinfo  {journal} {Sci. Adv.}\ }\textbf {\bibinfo {volume}
  {8}},\ \bibinfo {pages} {eadd3854} (\bibinfo {year} {2022})}\BibitemShut
  {NoStop}%
\bibitem [{\citenamefont {Bongs}\ \emph {et~al.}(2019)\citenamefont {Bongs},
  \citenamefont {Holynski}, \citenamefont {Vovrosh}, \citenamefont {Bouyer},
  \citenamefont {Condon}, \citenamefont {Rasel}, \citenamefont {Schubert},
  \citenamefont {Schleich},\ and\ \citenamefont {Roura}}]{bongs2019taking}%
  \BibitemOpen
  \bibfield  {author} {\bibinfo {author} {\bibfnamefont {K.}~\bibnamefont
  {Bongs}}, \bibinfo {author} {\bibfnamefont {M.}~\bibnamefont {Holynski}},
  \bibinfo {author} {\bibfnamefont {J.}~\bibnamefont {Vovrosh}}, \bibinfo
  {author} {\bibfnamefont {P.}~\bibnamefont {Bouyer}}, \bibinfo {author}
  {\bibfnamefont {G.}~\bibnamefont {Condon}}, \bibinfo {author} {\bibfnamefont
  {E.}~\bibnamefont {Rasel}}, \bibinfo {author} {\bibfnamefont
  {C.}~\bibnamefont {Schubert}}, \bibinfo {author} {\bibfnamefont {W.~P.}\
  \bibnamefont {Schleich}},\ and\ \bibinfo {author} {\bibfnamefont
  {A.}~\bibnamefont {Roura}},\ }\bibfield  {title} {\bibinfo {title} {Taking
  atom interferometric quantum sensors from the laboratory to real-world
  applications},\ }\href
  {https://doi.org/https://doi.org/10.1038/s42254-019-0117-4} {\bibfield
  {journal} {\bibinfo  {journal} {Nat. Rev. Phys.}\ }\textbf {\bibinfo {volume}
  {1}},\ \bibinfo {pages} {731} (\bibinfo {year} {2019})}\BibitemShut {NoStop}%
\bibitem [{\citenamefont {Sato}\ \emph {et~al.}(2025)\citenamefont {Sato},
  \citenamefont {Nishimura}, \citenamefont {Kaku}, \citenamefont {Otabe},
  \citenamefont {Kawasaki}, \citenamefont {Hosoya},\ and\ \citenamefont
  {Kozuma}}]{sato2025closed}%
  \BibitemOpen
  \bibfield  {author} {\bibinfo {author} {\bibfnamefont {T.}~\bibnamefont
  {Sato}}, \bibinfo {author} {\bibfnamefont {N.}~\bibnamefont {Nishimura}},
  \bibinfo {author} {\bibfnamefont {N.}~\bibnamefont {Kaku}}, \bibinfo {author}
  {\bibfnamefont {S.}~\bibnamefont {Otabe}}, \bibinfo {author} {\bibfnamefont
  {T.}~\bibnamefont {Kawasaki}}, \bibinfo {author} {\bibfnamefont
  {T.}~\bibnamefont {Hosoya}},\ and\ \bibinfo {author} {\bibfnamefont
  {M.}~\bibnamefont {Kozuma}},\ }\bibfield  {title} {\bibinfo {title}
  {Closed-loop measurements in an atom-interferometer gyroscope with
  compensation for velocity-dependent phase dispersion},\ }\href
  {https://doi.org/https://doi.org/10.1103/PhysRevApplied.23.044001} {\bibfield
   {journal} {\bibinfo  {journal} {Phys. Rev. Appl.}\ }\textbf {\bibinfo
  {volume} {23}},\ \bibinfo {pages} {044001} (\bibinfo {year}
  {2025})}\BibitemShut {NoStop}%
\bibitem [{\citenamefont {Cheiney}\ \emph {et~al.}(2018)\citenamefont
  {Cheiney}, \citenamefont {Fouch{\'e}}, \citenamefont {Templier},
  \citenamefont {Napolitano}, \citenamefont {Battelier}, \citenamefont
  {Bouyer},\ and\ \citenamefont {Barrett}}]{cheiney2018navigation}%
  \BibitemOpen
  \bibfield  {author} {\bibinfo {author} {\bibfnamefont {P.}~\bibnamefont
  {Cheiney}}, \bibinfo {author} {\bibfnamefont {L.}~\bibnamefont {Fouch{\'e}}},
  \bibinfo {author} {\bibfnamefont {S.}~\bibnamefont {Templier}}, \bibinfo
  {author} {\bibfnamefont {F.}~\bibnamefont {Napolitano}}, \bibinfo {author}
  {\bibfnamefont {B.}~\bibnamefont {Battelier}}, \bibinfo {author}
  {\bibfnamefont {P.}~\bibnamefont {Bouyer}},\ and\ \bibinfo {author}
  {\bibfnamefont {B.}~\bibnamefont {Barrett}},\ }\bibfield  {title} {\bibinfo
  {title} {Navigation-compatible hybrid quantum accelerometer using a kalman
  filter},\ }\href
  {https://doi.org/https://doi.org/10.1103/PhysRevApplied.10.034030} {\bibfield
   {journal} {\bibinfo  {journal} {Physical Review Applied}\ }\textbf {\bibinfo
  {volume} {10}},\ \bibinfo {pages} {034030} (\bibinfo {year}
  {2018})}\BibitemShut {NoStop}%
\bibitem [{\citenamefont {Lautier}\ \emph {et~al.}(2014)\citenamefont
  {Lautier}, \citenamefont {Volodimer}, \citenamefont {Hardin}, \citenamefont
  {Merlet}, \citenamefont {Lours}, \citenamefont {Pereira Dos~Santos},\ and\
  \citenamefont {Landragin}}]{lautier2014hybridizing}%
  \BibitemOpen
  \bibfield  {author} {\bibinfo {author} {\bibfnamefont {J.}~\bibnamefont
  {Lautier}}, \bibinfo {author} {\bibfnamefont {L.}~\bibnamefont {Volodimer}},
  \bibinfo {author} {\bibfnamefont {T.}~\bibnamefont {Hardin}}, \bibinfo
  {author} {\bibfnamefont {S.}~\bibnamefont {Merlet}}, \bibinfo {author}
  {\bibfnamefont {M.}~\bibnamefont {Lours}}, \bibinfo {author} {\bibfnamefont
  {F.}~\bibnamefont {Pereira Dos~Santos}},\ and\ \bibinfo {author}
  {\bibfnamefont {A.}~\bibnamefont {Landragin}},\ }\bibfield  {title} {\bibinfo
  {title} {Hybridizing matter-wave and classical accelerometers},\ }\href
  {https://doi.org/https://doi.org/10.1063/1.4897358} {\bibfield  {journal}
  {\bibinfo  {journal} {Appl. Phys. Lett.}\ }\textbf {\bibinfo {volume}
  {105}},\ \bibinfo {pages} {144102} (\bibinfo {year} {2014})}\BibitemShut
  {NoStop}%
\bibitem [{\citenamefont {Merlet}\ \emph {et~al.}(2009)\citenamefont {Merlet},
  \citenamefont {Le~Gou{\"e}t}, \citenamefont {Bodart}, \citenamefont
  {Clairon}, \citenamefont {Landragin}, \citenamefont {Dos~Santos},\ and\
  \citenamefont {Rouchon}}]{merlet2009operating}%
  \BibitemOpen
  \bibfield  {author} {\bibinfo {author} {\bibfnamefont {S.}~\bibnamefont
  {Merlet}}, \bibinfo {author} {\bibfnamefont {J.}~\bibnamefont
  {Le~Gou{\"e}t}}, \bibinfo {author} {\bibfnamefont {Q.}~\bibnamefont
  {Bodart}}, \bibinfo {author} {\bibfnamefont {A.}~\bibnamefont {Clairon}},
  \bibinfo {author} {\bibfnamefont {A.}~\bibnamefont {Landragin}}, \bibinfo
  {author} {\bibfnamefont {F.~P.}\ \bibnamefont {Dos~Santos}},\ and\ \bibinfo
  {author} {\bibfnamefont {P.}~\bibnamefont {Rouchon}},\ }\bibfield  {title}
  {\bibinfo {title} {Operating an atom interferometer beyond its linear
  range},\ }\href {https://doi.org/10.1088/0026-1394/46/1/011} {\bibfield
  {journal} {\bibinfo  {journal} {Metrologia}\ }\textbf {\bibinfo {volume}
  {46}},\ \bibinfo {pages} {87} (\bibinfo {year} {2009})}\BibitemShut {NoStop}%
\bibitem [{\citenamefont {Jekeli}(2005)}]{jekeli2005navigation}%
  \BibitemOpen
  \bibfield  {author} {\bibinfo {author} {\bibfnamefont {C.}~\bibnamefont
  {Jekeli}},\ }\bibfield  {title} {\bibinfo {title} {Navigation error analysis
  of atom interferometer inertial sensor},\ }\href
  {https://doi.org/https://doi.org/10.1002/j.2161-4296.2005.tb01726.x}
  {\bibfield  {journal} {\bibinfo  {journal} {Navigation}\ }\textbf {\bibinfo
  {volume} {52}},\ \bibinfo {pages} {1} (\bibinfo {year} {2005})}\BibitemShut
  {NoStop}%
\bibitem [{\citenamefont {Narducci}\ \emph {et~al.}(2022)\citenamefont
  {Narducci}, \citenamefont {Black},\ and\ \citenamefont
  {Burke}}]{narducci2022advances}%
  \BibitemOpen
  \bibfield  {author} {\bibinfo {author} {\bibfnamefont {F.~A.}\ \bibnamefont
  {Narducci}}, \bibinfo {author} {\bibfnamefont {A.~T.}\ \bibnamefont
  {Black}},\ and\ \bibinfo {author} {\bibfnamefont {J.~H.}\ \bibnamefont
  {Burke}},\ }\bibfield  {title} {\bibinfo {title} {Advances toward fieldable
  atom interferometers},\ }\href
  {https://doi.org/https://doi.org/10.1080/23746149.2021.1946426} {\bibfield
  {journal} {\bibinfo  {journal} {Advances in Physics: X}\ }\textbf {\bibinfo
  {volume} {7}},\ \bibinfo {pages} {1946426} (\bibinfo {year}
  {2022})}\BibitemShut {NoStop}%
\bibitem [{\citenamefont {Bonnin}\ \emph {et~al.}(2018)\citenamefont {Bonnin},
  \citenamefont {Diboune}, \citenamefont {Zahzam}, \citenamefont {Bidel},
  \citenamefont {Cadoret},\ and\ \citenamefont {Bresson}}]{bonnin2018a}%
  \BibitemOpen
  \bibfield  {author} {\bibinfo {author} {\bibfnamefont {A.}~\bibnamefont
  {Bonnin}}, \bibinfo {author} {\bibfnamefont {C.}~\bibnamefont {Diboune}},
  \bibinfo {author} {\bibfnamefont {N.}~\bibnamefont {Zahzam}}, \bibinfo
  {author} {\bibfnamefont {Y.}~\bibnamefont {Bidel}}, \bibinfo {author}
  {\bibfnamefont {M.}~\bibnamefont {Cadoret}},\ and\ \bibinfo {author}
  {\bibfnamefont {A.}~\bibnamefont {Bresson}},\ }\bibfield  {title} {\bibinfo
  {title} {New concepts of inertial measurements with multi-species atom
  interferometry},\ }\href
  {https://doi.org/https://doi.org/10.1007/s00340-018-7051-5} {\bibfield
  {journal} {\bibinfo  {journal} {Appl. Phys. B}\ }\textbf {\bibinfo {volume}
  {124}},\ \bibinfo {pages} {181} (\bibinfo {year} {2018})}\BibitemShut
  {NoStop}%
\bibitem [{\citenamefont {Avinadav}\ \emph {et~al.}(2020)\citenamefont
  {Avinadav}, \citenamefont {Yankelev}, \citenamefont {Firstenberg},\ and\
  \citenamefont {Davidson}}]{avinadav2020}%
  \BibitemOpen
  \bibfield  {author} {\bibinfo {author} {\bibfnamefont {C.}~\bibnamefont
  {Avinadav}}, \bibinfo {author} {\bibfnamefont {D.}~\bibnamefont {Yankelev}},
  \bibinfo {author} {\bibfnamefont {O.}~\bibnamefont {Firstenberg}},\ and\
  \bibinfo {author} {\bibfnamefont {N.}~\bibnamefont {Davidson}},\ }\bibfield
  {title} {\bibinfo {title} {Composite-{{Fringe Atom Interferometry}} for
  {{High-Dynamic-Range Sensing}}},\ }\href
  {https://doi.org/https://doi.org/10.1103/PhysRevApplied.13.054053} {\bibfield
   {journal} {\bibinfo  {journal} {Phys. Rev. Appl.}\ }\textbf {\bibinfo
  {volume} {13}},\ \bibinfo {pages} {054053} (\bibinfo {year}
  {2020})}\BibitemShut {NoStop}%
\bibitem [{\citenamefont {Yankelev}\ \emph {et~al.}(2020)\citenamefont
  {Yankelev}, \citenamefont {Avinadav}, \citenamefont {Davidson},\ and\
  \citenamefont {Firstenberg}}]{yankelev2020a}%
  \BibitemOpen
  \bibfield  {author} {\bibinfo {author} {\bibfnamefont {D.}~\bibnamefont
  {Yankelev}}, \bibinfo {author} {\bibfnamefont {C.}~\bibnamefont {Avinadav}},
  \bibinfo {author} {\bibfnamefont {N.}~\bibnamefont {Davidson}},\ and\
  \bibinfo {author} {\bibfnamefont {O.}~\bibnamefont {Firstenberg}},\
  }\bibfield  {title} {\bibinfo {title} {Atom interferometry with thousand-fold
  increase in dynamic range},\ }\bibfield  {journal} {\bibinfo  {journal} {Sci.
  Adv.}\ }\href {https://doi.org/10.1126/sciadv.abd0650}
  {10.1126/sciadv.abd0650} (\bibinfo {year} {2020})\BibitemShut {NoStop}%
\bibitem [{\citenamefont {Black}\ and\ \citenamefont
  {Kwolek}(2023)}]{black2023velocity}%
  \BibitemOpen
  \bibfield  {author} {\bibinfo {author} {\bibfnamefont {A.~T.}\ \bibnamefont
  {Black}}\ and\ \bibinfo {author} {\bibfnamefont {J.~M.}\ \bibnamefont
  {Kwolek}},\ }\bibfield  {title} {\bibinfo {title} {Velocity-modulated atom
  interferometry with enhanced dynamic range},\ }in\ \href
  {https://doi.org/https://doi.org/10.1117/12.2657192} {\emph {\bibinfo
  {booktitle} {Quantum Sensing, Imaging, and Precision Metrology}}},\ Vol.\
  \bibinfo {volume} {12447}\ (\bibinfo {organization} {SPIE},\ \bibinfo {year}
  {2023})\ pp.\ \bibinfo {pages} {179--183}\BibitemShut {NoStop}%
\bibitem [{\citenamefont {Meng}\ \emph {et~al.}(2024)\citenamefont {Meng},
  \citenamefont {Yan}, \citenamefont {Wang}, \citenamefont {Li}, \citenamefont
  {Xue},\ and\ \citenamefont {Feng}}]{Meng2024}%
  \BibitemOpen
  \bibfield  {author} {\bibinfo {author} {\bibfnamefont {Z.-X.}\ \bibnamefont
  {Meng}}, \bibinfo {author} {\bibfnamefont {P.-Q.}\ \bibnamefont {Yan}},
  \bibinfo {author} {\bibfnamefont {S.-Z.}\ \bibnamefont {Wang}}, \bibinfo
  {author} {\bibfnamefont {X.-J.}\ \bibnamefont {Li}}, \bibinfo {author}
  {\bibfnamefont {H.-B.}\ \bibnamefont {Xue}},\ and\ \bibinfo {author}
  {\bibfnamefont {Y.-Y.}\ \bibnamefont {Feng}},\ }\bibfield  {title} {\bibinfo
  {title} {Closed-loop dual-atom-interferometer inertial sensor with continuous
  cold atomic beams},\ }\href
  {https://doi.org/https://doi.org/10.1103/PhysRevApplied.21.034050} {\bibfield
   {journal} {\bibinfo  {journal} {Phys. Rev. Appl.}\ }\textbf {\bibinfo
  {volume} {21}},\ \bibinfo {pages} {034050} (\bibinfo {year}
  {2024})}\BibitemShut {NoStop}%
\bibitem [{\citenamefont {Xue}\ \emph {et~al.}(2015)\citenamefont {Xue},
  \citenamefont {Feng}, \citenamefont {Chen}, \citenamefont {Wang},
  \citenamefont {Yan}, \citenamefont {Jiang},\ and\ \citenamefont
  {Zhou}}]{xue2015continuous}%
  \BibitemOpen
  \bibfield  {author} {\bibinfo {author} {\bibfnamefont {H.}~\bibnamefont
  {Xue}}, \bibinfo {author} {\bibfnamefont {Y.}~\bibnamefont {Feng}}, \bibinfo
  {author} {\bibfnamefont {S.}~\bibnamefont {Chen}}, \bibinfo {author}
  {\bibfnamefont {X.}~\bibnamefont {Wang}}, \bibinfo {author} {\bibfnamefont
  {X.}~\bibnamefont {Yan}}, \bibinfo {author} {\bibfnamefont {Z.}~\bibnamefont
  {Jiang}},\ and\ \bibinfo {author} {\bibfnamefont {Z.}~\bibnamefont {Zhou}},\
  }\bibfield  {title} {\bibinfo {title} {A continuous cold atomic beam
  interferometer},\ }\href {https://doi.org/https://doi.org/10.1063/1.4913711}
  {\bibfield  {journal} {\bibinfo  {journal} {J. Appl. Phys.}\ }\textbf
  {\bibinfo {volume} {117}},\ \bibinfo {pages} {094901} (\bibinfo {year}
  {2015})}\BibitemShut {NoStop}%
\bibitem [{\citenamefont {Kwolek}\ \emph {et~al.}(2020)\citenamefont {Kwolek},
  \citenamefont {Fancher}, \citenamefont {Bashkansky},\ and\ \citenamefont
  {Black}}]{kwolek2020three}%
  \BibitemOpen
  \bibfield  {author} {\bibinfo {author} {\bibfnamefont {J.}~\bibnamefont
  {Kwolek}}, \bibinfo {author} {\bibfnamefont {C.}~\bibnamefont {Fancher}},
  \bibinfo {author} {\bibfnamefont {M.}~\bibnamefont {Bashkansky}},\ and\
  \bibinfo {author} {\bibfnamefont {A.}~\bibnamefont {Black}},\ }\bibfield
  {title} {\bibinfo {title} {Three-dimensional cooling of an atom-beam source
  for high-contrast atom interferometry},\ }\href
  {https://doi.org/https://doi.org/10.1103/PhysRevApplied.13.044057} {\bibfield
   {journal} {\bibinfo  {journal} {Phys. Rev. Appl.}\ }\textbf {\bibinfo
  {volume} {13}},\ \bibinfo {pages} {044057} (\bibinfo {year}
  {2020})}\BibitemShut {NoStop}%
\bibitem [{\citenamefont {Kwolek}\ and\ \citenamefont
  {Black}(2022)}]{kwolek2022continuous}%
  \BibitemOpen
  \bibfield  {author} {\bibinfo {author} {\bibfnamefont {J.}~\bibnamefont
  {Kwolek}}\ and\ \bibinfo {author} {\bibfnamefont {A.}~\bibnamefont {Black}},\
  }\bibfield  {title} {\bibinfo {title} {Continuous sub-doppler-cooled atomic
  beam interferometer for inertial sensing},\ }\href
  {https://doi.org/https://doi.org/10.1103/PhysRevApplied.17.024061} {\bibfield
   {journal} {\bibinfo  {journal} {Phys. Rev. Appl.}\ }\textbf {\bibinfo
  {volume} {17}},\ \bibinfo {pages} {024061} (\bibinfo {year}
  {2022})}\BibitemShut {NoStop}%
\bibitem [{\citenamefont {d’Armagnac~de Castanet}\ \emph
  {et~al.}(2024)\citenamefont {d’Armagnac~de Castanet}, \citenamefont
  {Des~Cognets}, \citenamefont {Arguel}, \citenamefont {Templier},
  \citenamefont {Jarlaud}, \citenamefont {M{\'e}noret}, \citenamefont
  {Desruelle}, \citenamefont {Bouyer},\ and\ \citenamefont
  {Battelier}}]{d2024atom}%
  \BibitemOpen
  \bibfield  {author} {\bibinfo {author} {\bibfnamefont {Q.}~\bibnamefont
  {d’Armagnac~de Castanet}}, \bibinfo {author} {\bibfnamefont
  {C.}~\bibnamefont {Des~Cognets}}, \bibinfo {author} {\bibfnamefont
  {R.}~\bibnamefont {Arguel}}, \bibinfo {author} {\bibfnamefont
  {S.}~\bibnamefont {Templier}}, \bibinfo {author} {\bibfnamefont
  {V.}~\bibnamefont {Jarlaud}}, \bibinfo {author} {\bibfnamefont
  {V.}~\bibnamefont {M{\'e}noret}}, \bibinfo {author} {\bibfnamefont
  {B.}~\bibnamefont {Desruelle}}, \bibinfo {author} {\bibfnamefont
  {P.}~\bibnamefont {Bouyer}},\ and\ \bibinfo {author} {\bibfnamefont
  {B.}~\bibnamefont {Battelier}},\ }\bibfield  {title} {\bibinfo {title} {Atom
  interferometry at arbitrary orientations and rotation rates},\ }\href
  {https://doi.org/https://doi.org/10.1038/s41467-024-50804-0} {\bibfield
  {journal} {\bibinfo  {journal} {Nat. Commun.}\ }\textbf {\bibinfo {volume}
  {15}},\ \bibinfo {pages} {6406} (\bibinfo {year} {2024})}\BibitemShut
  {NoStop}%
\bibitem [{\citenamefont {Kasevich}\ and\ \citenamefont
  {Chu}(1991)}]{kasevich1991atomic}%
  \BibitemOpen
  \bibfield  {author} {\bibinfo {author} {\bibfnamefont {M.}~\bibnamefont
  {Kasevich}}\ and\ \bibinfo {author} {\bibfnamefont {S.}~\bibnamefont {Chu}},\
  }\bibfield  {title} {\bibinfo {title} {Atomic interferometry using stimulated
  raman transitions},\ }\href
  {https://doi.org/https://doi.org/10.1103/PhysRevLett.67.181} {\bibfield
  {journal} {\bibinfo  {journal} {Phys. Rev. Lett.}\ }\textbf {\bibinfo
  {volume} {67}},\ \bibinfo {pages} {181} (\bibinfo {year} {1991})}\BibitemShut
  {NoStop}%
\bibitem [{\citenamefont {Here}()}]{Detail1}%
  \BibitemOpen
  \bibfield  {author} {\bibinfo {author} {\bibnamefont {Here}},\ }\bibinfo
  {note} {the interferometric phase is given by
  $\Phi_{\mathrm{MZ}}=\phi_{1}-2\phi_{2}+\phi_{3}$, where $\phi_i$ $(i=1,2,3)$
  denote the phases of the three Raman pulses encoded by the atoms. The phase
  contribution of the left interferometer can be written as
  $\phi_{MZ,1}=k_{\mathrm{eff}}aT^2 -2k_{\mathrm{eff}}\Omega LT -2\pi\delta f
  t_0 -2[2\pi f_r(t_0+T)] +2\pi\delta f(t_0+2T) +\phi_0$, where $\phi_0$
  represents the initial phase of the interferometer. The phase evolution of
  the right interferometer can be obtained in an analogous manner.}\BibitemShut
  {Stop}%
\bibitem [{\citenamefont {Yan}\ \emph {et~al.}(2025)\citenamefont {Yan},
  \citenamefont {Jia}, \citenamefont {Shen}, \citenamefont {Xin},\ and\
  \citenamefont {Feng}}]{Yan2025}%
  \BibitemOpen
  \bibfield  {author} {\bibinfo {author} {\bibfnamefont {P.-Q.}\ \bibnamefont
  {Yan}}, \bibinfo {author} {\bibfnamefont {W.-C.}\ \bibnamefont {Jia}},
  \bibinfo {author} {\bibfnamefont {K.}~\bibnamefont {Shen}}, \bibinfo {author}
  {\bibfnamefont {Y.}~\bibnamefont {Xin}},\ and\ \bibinfo {author}
  {\bibfnamefont {Y.-Y.}\ \bibnamefont {Feng}},\ }\bibfield  {title} {\bibinfo
  {title} {A continuous dual-axis atomic interferometric inertial sensor},\
  }\href {https://doi.org/10.1109/TIM.2025.3644533} {\bibfield  {journal}
  {\bibinfo  {journal} {IEEE Trans. Instrum. Meas.}\ }\textbf {\bibinfo
  {volume} {74}},\ \bibinfo {pages} {1} (\bibinfo {year} {2025})}\BibitemShut
  {NoStop}%
\bibitem [{Det()}]{Detail2}%
  \BibitemOpen
  \bibinfo {note} {A factor $\alpha(v)$ is introduced to characterize the
  mismatch between the effective velocities $v_{eq,r}$ and $v_{eq,f}$. The term
  $\Phi_+$ can then be expressed as $\Phi_+(v)=(-2k_{\mathrm{eff}}\Omega
  L^2/\alpha(v)-4\pi f_r L)/v$. When $\alpha(v)$ is not constant for each
  velocity $v$, the closed-loop law $\Omega=K f_r$ no longer cancels the
  velocity dependence $\partial^{n}\Phi_+/\partial v^{n}$ , which results in
  residual velocity-induced dephasing.}\BibitemShut {Stop}%
\end{thebibliography}%
\end{document}